\documentclass[letterpaper,twocolumn,english,prl, aps, showpacs]{revtex4}
\usepackage[T1]{fontenc}
\usepackage[latin9]{inputenc}
\usepackage{amsmath}
\usepackage{graphicx}
\usepackage{amssymb}

\usepackage{psfrag}

\usepackage{babel}

\begin{document}

\title{From Networks of Unstable Attractors to Heteroclinic Switching}

\author{Christoph Kirst$^{1-4}$ and Marc Timme$^{1,2}$}

\address{$^{1}$Network Dynamics Group, Max Planck Institute for Dynamics
and Self-Organization (MPIDS) and\\
$^{2}$Bernstein Center for Computational Neuroscience (BCCN) Göttingen,
37073 Göttingen, Germany\\
$^{3}$Fakultät für Physik, Georg-August-Universität Göttingen,
Germany\\
$^{4}$DAMTP, Centre for Mathematical Sciences, Cambridge University,
Cambridge CB3 0WA, UK}
\begin{abstract}
We present a dynamical system that naturally exhibits two unstable
attractors that are completely enclosed by each others basin volume.
This counter-intuitive phenomenon occurs in networks of pulse-coupled
oscillators with delayed interactions. We analytically show that upon
continuously removing a local non-invertibility of the system, the
two unstable attractors become a set of two non-attracting saddle
states that are heteroclinically connected. This transition equally
occurs from larger networks of unstable attractors to heteroclinic
structures and constitutes a new type of singular bifurcation in dynamical
systems.
\end{abstract}


\maketitle
The concepts of attractor and stability are at the core of dynamical
systems theory \cite{Katok} because attractivity and stability determine
the long term behavior and often the typical properties of a system.
 Attraction and stability which may change via bifurcations are thus
fundamental to modeling in all of science and engineering. For systems
with smooth and invertible flows these concepts have long been studied
and are well understood, allowing classifications of dynamical systems
and their bifurcations, for example by using topological equivalence
and normal forms. 

Dynamical systems with non-smooth or non-invertible flows, such as
hybrid or Fillipov systems \cite{Hybrid}, are far less understood
although they model a variety of natural phenomena, ranging from the
mechanics of stick-slip motion and the switching dynamics of electrical
circuits, to the generation of earthquakes and the spiking activity
of neural networks \cite{Examples,Timme,UnstableAttr}. For instance,
spiking neurons interact by sending and receiving electrical pulses
at discrete instances of time that interrupt the intermediate smooth
interaction-free dynamics. This neural dynamics and similarly that
of, e.g., cardiac pacemaker cells, plate tectonics in earthquakes,
chirping crickets and flashing fireflies are often modeled as pulse-coupled
oscillators.

Such hybrid systems display dynamics very different from that of temporally
continuous or temporally discrete systems. Networks of oscillators
with global homogeneous delayed pulse-coupling may robustly exhibit
\emph{unstable attractors} \cite{UnstableAttr} (invariant periodic
orbits that are Milnor attractors \cite{Katok} but locally unstable).
In the presence of noise, these systems exhibit a dynamics akin to
heteroclinic switching \cite{Switching1}, a feature that is functionally
relevant in many natural systems such as in neural, weather and population
dynamics \cite{Switching1,Switching2}. Rigorous analysis \cite{Ashwin}
shows that invertible systems in general cannot have unstable attractors
and that a saddle state can in principle be converted to an unstable
attractor by locally adding a non-invertible dynamics onto the stable
manifold. However, the potential relation of unstable attractors to
heteroclinic cycles is not well understood and it is unknown whether
and how unstable attractors may be created or destroyed via bifurcations.

In a network of pulse-coupled oscillators we here demonstrate the
existence of two unstable attractors that are enclosed by the basin
of attraction of each other. We explain this counter-intuitive phenomenon:
Continuously lifting the local non-invertibility of the system with
two unstable attractors creates a standard heteroclinic two-cycle.
This transition equally occurs from large networks of unstable attractors
to heteroclinic structures and constitutes a new type of singular
bifurcation in hybrid dynamical systems. 

We consider a network of $N$ oscillatory units with a state defined
by a phase-like variable $\phi_{i}(t)\in\mathbb{R}$, $i\in\left\{ 1,2,\dots,N\right\} $,
that increases uniformly in time $t$ \begin{equation}
\frac{d}{dt}\phi_{i}=1\,.\label{eq: model}\end{equation}
Upon crossing a threshold at time $t_{s}$, $\phi_{i}\left(t_{s}\right)\ge1$,
unit $i$ is instantaneously reset,\begin{equation}
\phi_{i}\left(t_{s}^{+}\right):=\lim_{r\searrow0}\phi_{i}\left(t_{s}+r\right)=K\left(\phi_{i}\left(t_{s}\right)\right)\:.\label{eq: reset}\end{equation}
Here $K\left(\phi\right)=U^{-1}\left(R\left(U\left(\phi\right)-1\right)\right)$
is determined by a smooth, unbounded, strictly monotonic increasing
\emph{rise function} $U\left(\phi\right)$ normalized to $U\left(0\right)=0$
and $U\left(1\right)=1$ and a smooth non-negative \emph{reset function}
$R$ satisfying $R(0)=0$. In addition to the reset \eqref{eq: reset}
a pulse is sent which is received by all units $j$ after a delay
time $\tau>0$, inducing a phase jump\begin{equation}
\phi_{j}\left(t_{s}+\tau\right)=H_{\varepsilon_{ji}}\left(\phi_{j}\left(\left(t_{s}+\tau\right)^{-}\right)\right)\label{eq:phiJump}\end{equation}
with \emph{interaction function} $H_{\varepsilon}\left(\phi\right)=U^{-1}\left(U\left(\phi\right)+\varepsilon\right)$
and coupling strength $\varepsilon_{ji}$ from unit $i$ to unit $j$.
We set $J_{\varepsilon}\left(\phi\right)=K\circ H_{\varepsilon}\left(\phi\right)$
and denote a phase shift by $S_{\eta}(\phi)=\phi+\eta$.

\begin{figure}
\begin{centering}
\includegraphics{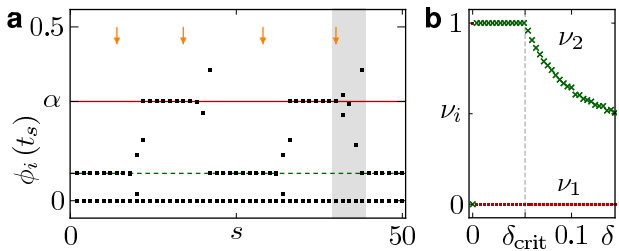}
\par\end{centering}

\caption{(color online) \label{fig:Two-interwoven-unstable}Two unstable attractors
enclosed by the basins of each other ($c=0$). \textbf{(a)} Phases
$\phi_{i}\left(t_{s}\right)$ (dots) of all units at times $t_{s}$
just after the $s$-th reset of a reference unit $i=1$. Lines indicate
the phases on the invariant orbit $A_{1}$ ($\alpha$, solid) and
$A_{2}$ (dashed). Arrows mark times of small phase perturbations
which induce switches from $A_{1}$ to $A_{2}$ or vice versa. The
shaded area highlights a switch from $A_{1}$ to $A_{2}$ that is
shown in detail in Fig.~\ref{fig:Reduced-phase-space}. \textbf{(b)}
Fraction $\nu_{i}$ of 5000 trajectories reaching the periodic orbit
$A_{i}$ ($\bullet:i=1$, $\times:i=2$) starting from random phases
distributed uniformly in a box of side width $2\delta$ centered around
$\phi=a_{1}$ on the orbit $A_{1}$. For $0<\delta<\delta_{\mathrm{crit}}\approx0.05$
all trajectories reach the orbit $A_{2}$ ($\nu_{2}=1$), indicating
that $A_{1}$ is enclosed by the basin volume of $A_{2}$ and in
particular that $A_{1}$ is an unstable attractor. }

\end{figure}

This system represents, for instance, an abstract model of neuronal
oscillators with a membrane potential $u_{i}\left(t\right)=U\left(\phi_{i}\left(t\right)\right)$.
The neurons' responses to inputs are described by increasing the potentials
instantaneously by an amount $\varepsilon_{ij}$, that represents
the transferred charge from the pulse sending (presynaptic) neuron
$j$ to the pulse receiving (postsynaptic) one $i$. If this input
is supra-threshold, $u_{i}\left(t\right)=u_{i}\left(t^{-}\right)+\varepsilon_{ij}>1$,
unit $i$ is \emph{partially reset} to \begin{equation}
u_{i}\left(t^{+}\right)=R\left(u_{i}\left(t\right)-1\right)\geq0\,.\label{eq:PartialResetU}\end{equation}
 This accounts for remaining synaptic input charges which are not
used to reach the threshold and which contribute to the potential
after reset \cite{Kirst}. For $R\left(\zeta\right)\equiv0$ we recover
the model analyzed in previous studies \cite{UnstableAttr} which
has a local non-invertibility since the original phase of a unit cannot
be recovered after it received supra-threshold input and was reset
to $J_{\varepsilon}\left(\phi\right)\equiv0$. For an invertible $R$
the flow becomes locally time invertible.

Here we focus on a homogeneous network of all-to-all coupled excitatory
units without self-interaction, i.e. $\varepsilon_{ij}=\left(1-\delta_{ij}\right)$$\varepsilon$,
$\varepsilon>0$. The permutation symmetry implies invariant subspaces
of two or more synchronized units and thus the possibility of robust
heteroclinic cycles, cf. \cite{Krupa}. For the numerical simulations
presented below we fix $\varepsilon=0.23$, $\tau=0.02$, a rise function
$U(\phi)=\frac{1}{b}\log\left(1+\left(\exp(b)-1\right)\phi\right)$
with $b=4.2$ and partial reset $R\left(\zeta\right)=c\zeta$ with
parameter $c\in\left[0,1\right]$ which is invertible for all $c>0$.
For these parameters the model exhibits short switching times between
periodic orbits which simplifies the presentation of the analysis
below; however, the studied phenomena is robust against structural
perturbations in $\tau$, $\varepsilon$ and the function $U$.

For locally non-invertible dynamics ($c=0$) the above system exhibits
unstable attractors in a large fraction of parameter space and for
different network sizes $N$ \cite{UnstableAttr,Ashwin}. For the
above parameters, the smallest system in which we observed unstable
attractors has $N=4$ units. Curiously, numerical simulations, e.g.
Fig.~\ref{fig:Two-interwoven-unstable}a, indicate that such a system
exhibits two unstable attractors each of which is fully enclosed by
the basin volume of the other attractor, Fig.~\ref{fig:Two-interwoven-unstable}b. 

We confirm these numerical findings analytically. Given a periodic
orbit $A$, define the basin of attraction $\mathcal{B}(A)$ as the
set of points in state space that converge to $A$ in the long time
limit. Below we show that in the system \eqref{eq: model}-\eqref{eq:phiJump}
with $R(\zeta)\equiv0$ there is a pair of periodic orbits $A_{1}$
and $A_{2}$  such that  a full measure set of points of an open
neighborhood of $A_{1}$ is contained in the basin $\mathcal{B}(A_{2})$
and vice versa. 

\begin{figure}
\begin{centering}
\includegraphics{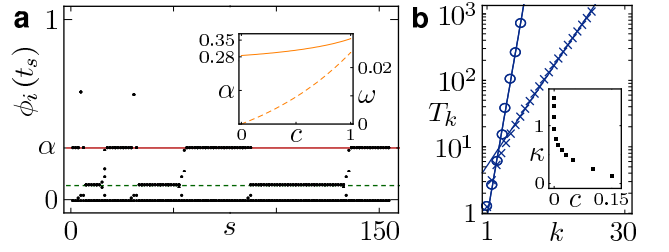}
\par\end{centering}

\caption{(color online) \label{fig:Heteroclinic-Switching}Heteroclinic switching
($c>0$). \textbf{(a)} Phases $\phi_{i}\left(t_{s}\right)$ (dots)
as in Fig.~\ref{fig:Two-interwoven-unstable}a for $c=0.05$. The
invariant periodic orbits $A_{i}$, being unstable attractors at $c=0$,
still exist for $c>0$ (solid and dashed line). Starting in a state
near $A_{1}$ leads to repeated switching between the two states.
Inset: Phase difference $\alpha$ \eqref{eq:alphaImplicit} (solid)
and side width \textbf{$w$} of the set \textbf{$\mathcal{D}_{i}'$}
\eqref{eq:width} (dashed) change continuously upon increasing $c$
from zero. \textbf{(b)} Switching times $T_{k}$ to the $k$-th switch
($\times$: $c=0.1$, $\circ$: $c=0.01$) increase exponentially
with $k$, indicating that the dynamics evolve near a heteroclinic
cycle between the invariant states. Inset: Fitting $T_{k}=\gamma e^{\kappa k}$
to the switching times for several values of $c$ we find a divergence
of $\kappa$ as $c\rightarrow0$.}

\end{figure}

To study the dynamics in detail we use an event based analysis, cf.
e.g.~\cite{Timme}. The event when a unit $i$ sends a pulse is denoted
by $s_{i}$, the reception of a pulse from unit $j$ by $r_{j}$ and
simultaneous events are enclosed in parentheses. For given parameter
$c\in[0,1]$, a simple saddle periodic orbit $A_{1}$ (cf. Figs.~\ref{fig:Two-interwoven-unstable}a
and \ref{fig:Heteroclinic-Switching}a) is uniquely determined by
the cyclic event sequence \begin{equation}
E\left(A_{1}\right)=\left(s_{1},s_{2}\right)\left(r_{1},r_{2},s_{3},s_{4}\right)\left(r_{3},r_{4}\right)\label{eq: A1}\end{equation}
By exchanging the indices $(1,2)\leftrightarrow(3,4)$ in \eqref{eq: A1}
we obtain the event sequence of a permutation equivalent periodic
orbit $A_{2}$. Both orbits lie in the intersection of the two invariant
subspaces $\left\{ \phi_{1}=\phi_{2}\right\} $ and $\left\{ \phi_{3}=\phi_{4}\right\} $
with synchronized units $(1,2)$ and $(3,4)$, respectively, allowing
a robust heteroclinic connection between them. As it turns out below
the local stability and non-local attractivity properties of the
$A_{i}$ depend on the parameter $c$. 

We now first locally reduce the infinite dimensional state space of
the hybrid dynamical system with delayed coupling to three dimensions:
Local to $A_{1}$ and $A_{2}$ the state space reduces in finite time
\cite{Ashwin} to an eight-dimensional state space spanned by the
four phases $\mathbf{\phi}=\left(\phi_{1},\phi_{2},\phi_{3},\phi_{4}\right)$,
and the four times $\sigma_{i}\ge0$, $i\in\left\{ 1,\dots,4\right\} $
elapsed since the most recent pulse generation of oscillator $i$.
We consider the subset $\mathcal{M}=\left\{ \left(\phi,\sigma\right)\,|\,\sigma_{i}>\tau,\: i\in\left\{ 1,\dots,4\right\} \right\} $
of the state space where all pulses have been received: Then the state
space is effectively four dimensional, since the exact values of the
$\sigma_{i}>\tau$ do not influence the dynamics. Due to the uniform
phase shift \eqref{eq: model}, $A_{1}$ is a straight line in $\mathcal{M}$
after the last and before the first event in the sequence \eqref{eq: A1}.
We denote the point in the center of this line by $a_{1}$ and consider
states with phases $\mathbf{\phi}=a_{1}+\left(\delta_{1},\delta_{2},\delta_{3},\delta_{4}\right)$
in a neighborhood. Because of shift invariance we may further fix
$\delta_{1}=0$, being left with a locally three-dimensional representation
$\mathcal{P}_{1}\subset\mathbb{R}^{3}$ of the original state space
with states $\left(\delta_{2},\delta_{3},\delta_{4}\right)\in\mathcal{P}_{1}$.
Similarly, we have a local three-dimensional representation $\left(\delta_{4},\delta_{1},\delta_{2}\right)\in\mathcal{P}_{2}$
of the state space around $a_{2}\in\mathcal{M}\cap A_{2}\,$, constructed
analogously to $a_{1}$, this time fixing $\delta_{3}=0$. There is
an open neighborhood of $A_{i}$ in the full eight-dimensional state
space from which every orbit crosses $\mathcal{P}_{i}$  after at
most eight events (one cycle). In this sense $\mathcal{P}_{i}$ is
a three-dimensional Poincar\'{e} section in a neighborhood of $A_{i}$. 

\begin{figure}
\begin{centering}
\includegraphics{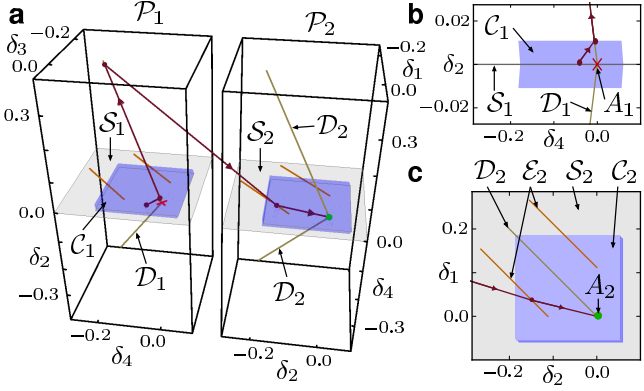}
\par\end{centering}

\caption{(color online) \label{fig:Reduced-phase-space}Structure of the three-dimensional
reduced state space for $c=0$ showing that the $A_{i}$ are unstable
attractors enclosed by the basins of each other. \textbf{(a)} Representations
$\mathcal{P}_{i}$ of the state space in a neighborhood of $A_{1}\in\mathcal{P}_{1}$
(cross) and $A_{2}\in\mathcal{P}_{2}$ (ball).  All trajectories
starting in the set $\mathcal{C}_{1}$ (close to $A_{1}$) lead to
a switch to $A_{2}$. The line with arrows shows a sample trajectory
of the marked switch from $A_{1}$ to $A_{2}$ in Fig.~\ref{fig:Two-interwoven-unstable}a.\textbf{}
\textbf{(b)} Projection of $\mathcal{P}_{1}$ onto the $\delta_{2}$-$\delta_{4}$
plane and \textbf{(c)} of $\mathcal{P}_{2}$ onto the $\delta_{1}$-$\delta_{2}$
plane, illustrating that, except for the lower dimensional subset
$\mathcal{S}_{i}$, the attractor $A_{i}$ is enclosed by $\mathcal{C}_{i}$,
i.e. the $A_{i}$ are unstable attractors. }

\end{figure}

For arbitrary $c\in[0,1]$ there are regions in $\mathcal{P}_{1}$
and $\mathcal{P}_{2}$ from which all trajectories evolve back to
points in either $\mathcal{P}_{1}$ or $\mathcal{P}_{2}$. Between
these regions we derive return maps and their domains which follow
directly from the definition of the local state space and the event
sequence (Fig. \ref{fig:Reduced-phase-space} visualizes domains of
the key maps and a sample trajectory for $c=0$). For instance, the
orbit $A_{1}$ is enclosed by the three-dimensional domain $\mathcal{C}_{1}\subset\mathcal{P}_{1}$
of the map $F:\mathcal{C}_{1}\rightarrow\mathcal{P}_{1}$, \begin{align}
F\left(\delta_{2},\delta_{3},\delta_{4}\right) & =\mathrm{\big(}\mathrm{sign}\left(\delta_{2}\right)\big[H_{2\varepsilon}\circ S_{\tau}\circ H_{\varepsilon}\left(\tau+\left|\delta_{2}\right|\right)\nonumber \\
 & \,\,\,\,-H_{2\varepsilon}\circ S_{\tau+\left|\delta_{2}\right|}\circ H_{\varepsilon}\left(\tau-\left|\delta_{2}\right|\right)\big],\delta_{3}',\delta_{4}'\big)\label{eq: C11}\end{align}
which is determined by the event sequence \begin{equation}
E\left(\mathcal{C}_{1}\right)=\left(s_{1}\right)\left(s_{2}\right)\left(r_{1}\right)\left(r_{2},s_{3},s_{4}\right)\left(r_{3},r_{4}\right)\label{eq:sequenceC1}\end{equation}
or its equivalent with permuted indices $1\leftrightarrow2$. Here\begin{eqnarray}
\delta_{i}' & = & H_{\varepsilon}\circ S_{\tau}\circ J_{\varepsilon}\left(H_{\varepsilon}\left(\alpha+\tau+\delta_{i}\right)+\left|\delta_{2}\right|\right)\nonumber \\
 &  & \quad\quad\quad+1-H_{2\varepsilon}\circ S_{\tau}\circ H_{\varepsilon}\left(\tau+\left|\delta_{2}\right|\right)-\alpha\label{eq:34synch}\end{eqnarray}
 for $i\in\left\{ 3,4\right\} $, where the phase difference $\alpha$
between the two synchronized clusters at $a_{1}$ is determined by
\begin{equation}
\alpha=H_{\varepsilon}\circ S_{\tau}\circ J_{2\varepsilon}\left(\alpha+\tau\right)+1-H_{2\varepsilon}\circ S_{\tau}\circ H_{\varepsilon}\left(\tau\right).\label{eq:alphaImplicit}\end{equation}
For $\left|\delta_{2}\right|>0$, $F$ is expanding in the $\delta_{2}$-direction
since \begin{equation}
\left|\left(F\left(\mathbf{\delta}\right)\right)_{2}\right|>k\left|\delta_{2}\right|\label{eq: C11expansion}\end{equation}
 with $k=\min_{\phi\in[0,1]}H_{\varepsilon}'(\phi)>1$. For $\delta_{2}=0$
we obtain a map with the same explicit form as in \eqref{eq: C11}
but with event sequence as in \eqref{eq: A1} whose domain is a subset
of the two-dimensional invariant set $\mathcal{S}_{1}=\left\{ \delta\in\mathcal{P}_{1}\,|\:\delta_{2}=0\right\} $
where units $1$ and $2$ are synchronized. States in $\mathcal{S}_{1}$
converge to $A_{1}$ in the long time limit. Similarly, points in
$\mathcal{S}_{2}=\left\{ \delta\in\mathcal{P}_{2}\,|\:\delta_{4}=0\right\} $
reach $A_{2}$ asymptotically. 

If the system is locally non-invertible ($c=0$) the dynamics is as
follows (see Fig. \ref{fig:Reduced-phase-space}): Since $J_{\varepsilon}(\phi)\equiv0$,
$\delta_{3}'=\delta_{4}'$ according to \eqref{eq:34synch} and hence
$F$ maps $\mathcal{C}_{1}$ into two one-dimensional lines $\mathcal{D}_{1}=F\left(\mathcal{C}_{1}\right)$.
 Since $F$ is expanding in $\delta_{2}$ \eqref{eq: C11expansion},
all points in $\mathcal{D}_{1}\cap\mathcal{C}_{1}$ are mapped after
a finite number of interactions into $\mathcal{D}_{1}\setminus\mathcal{C}_{1}$.
The set $\mathcal{D}_{1}\setminus\mathcal{C}_{1}$ is mapped to $\mathcal{E}_{2}\subset\mathcal{S}_{2}$
and from there to the attractor $A_{2}$. In Fig.~\ref{fig:Reduced-phase-space}
we have plotted a sample trajectory for the switch marked in Fig.~\ref{fig:Two-interwoven-unstable}a.
For the positive measure set $\mathcal{C}_{1}\cup\mathcal{S}_{1}$,
that encloses $A_{1}$, we thus have $\mathcal{C}_{1}\subset\mathcal{B}\left(A_{2}\right)$
and only the zero measure subset $\mathcal{S}_{1}$ converges to $A_{1}$.
Thus $A_{1}$ is an unstable attractor. Permutation symmetry implies
analogous dynamics near $A_{2}$. Taken together, for $c=0$ the periodic
orbits $A_{1}$ and $A_{2}$ are unstable attractors enclosed by the
basins of each other.

If we remove the local non-invertibility ($c>0$), the dynamics changes
qualitatively as shown in Fig.~\ref{fig:Heteroclinic-Switching}:
The two periodic orbits $A_{i}$ with event sequence \eqref{eq: A1}
still exist, only the phase difference $\alpha$ changes continuously
with $c$ (Fig. \ref{fig:Heteroclinic-Switching}a). Starting in a
state near one of the $A_{i}$ leads to trajectories with switching
between both. The switching time increases exponentially with the
number of switches (Fig.~\ref{fig:Heteroclinic-Switching}b) indicating
that these dynamics originate from an orbit near a heteroclinic two-cycle.
Furthermore the switching times diverge as $c\rightarrow0$ (cf. Fig.~\ref{fig:Heteroclinic-Switching}b),
suggesting the transition to a network of unstable attractors at $c=0$.
Indeed, the structure of the domains of all return maps does not change
qualitatively when $c$ increases from zero. However, since $J_{\varepsilon}$
becomes invertible for $c>0$, according to (\ref{eq:34synch}) a
phase difference $\left|\delta_{3}-\delta_{4}\right|$ shrinks under
the return map $F$, but does not collapse to zero as for $c=0$;
hence the image $\mathcal{D}_{1}'=F\left(\mathcal{C}_{1}\right)$
stays three-dimensional. It consists of tubes (around the original
lines $\mathcal{D}_{1}$) with a square cross-section of side width
\begin{equation}
w(c)=H_{\varepsilon}\circ S_{\tau}\circ U^{-1}\left(c\varepsilon\right)-H_{\varepsilon}\circ S_{\tau}\circ U^{-1}\left(0\right)\label{eq:width}\end{equation}
that continuously increases with $c$ from $w(0)=0$ (Fig.~\ref{fig:Heteroclinic-Switching}a).
This reflects the local $c$-dependent contraction of the state space
according to (\ref{eq: C11}) and (\ref{eq:34synch}). All maps with
domains that have a non-empty intersection with $\mathcal{D}_{1}'$
map $\mathcal{D}_{1}'$ to a  three-dimensional state space volume
around $A_{2}$ that is a subset of $\mathcal{C}_{2}\cup\mathcal{S}_{2}$.
Taken together, states in the three-dimensional set $\mathcal{C}_{1}$
evolve to states in a positive measure subset of $\mathcal{C}_{2}\cup\mathcal{S}_{2}$
that encloses $A_{2}$. Using symmetry again, $\mathcal{C}_{2}$ is
analogously mapped to a subset of $\mathcal{C}_{1}\cup\mathcal{S}_{1}$.
This explains the observed switching. 

The unstable attractors are converted to non-attracting saddles by
locally removing the non-invertibility of the dynamics, which is reflected
in the expansion of $\mathcal{D}_{i}$, $i\in\left\{ 1,2\right\} $
to positive measure sets $\mathcal{D}_{i}'$ when increasing $c$
from zero. Moreover, states in the subset of $\mathcal{C}_{1}$ with
synchronized units $3$ and $4$, i.e. states in the set $\left\{ \delta\in\mathcal{C}_{1}\,|\,\delta_{3}=\delta_{4}\right\} $
are mapped to $\mathcal{S}_{2}$ and thus reach the orbit $A_{2}$
asymptotically. Hence, this set together with all its image points
in $\mathcal{P}_{1}$ and $\mathcal{P}_{2}$ form a heteroclinic connection
from $A_{1}$ to $A_{2}$. Thus, by symmetry, the network of two unstable
attractors ($c=0$) continuously bifurcates to a heteroclinic two-cycle
($c>0$). 

The underlying mechanism relies on the interplay of the local instability
\eqref{eq: C11expansion} and the parameter dependent contraction
induced by the reset \eqref{eq:PartialResetU}, implying the same
transition in larger systems (not shown). For locally non-invertible
dynamics these display larger networks of unstable attractors \cite{UnstableAttr}
with a link between two attractors $A_{i}\rightarrow A_{j}$ if every
neighborhood of $A_{i}$ contains a positive basin volume of $A_{j}$.
Upon lifting the local non-invertibility each link in this network
is replaced by a heteroclinic connection. 

In summary, we have presented and analyzed the counter-intuitive phenomenon
of two unstable attractors that are enclosed by each other's basin
volume. We explained this phenomenon by showing that there is a continuous
transition from two unstable attractors to a heteroclinic two-cycle.
Larger networks of unstable attractors equally show this transition
to more complex heteroclinic structures. It constitutes a new type
of singular bifurcation in dynamical systems and establishes the first
known bifurcation of unstable attractors. Moreover, our results show
that this bifurcation occurs upon continuously removing the non-invertibility
of the system, whereas both the non-invertible ($c=0$) and the locally
invertible ($c>0$) system exhibit equally discontinuous interactions.
This explicitly demonstrates that the local non-invertibility and
not the discontinuity is responsible for the creation of unstable
attractors.

The continuity of the bifurcation has theoretical and practical consequences:
For instance, one may investigate features of a system exhibiting
heteroclinic switching \cite{Switching1} by studying its limiting
counterpart with unstable attractors. Furthermore, this may help designing
systems with specific heteroclinic structure, for instance in artificial
neural networks, and guide our understanding of time series of switching
phenomena in nature, cf. \cite{Switching2}. The associated limiting
systems with unstable attractors may not only be analytically accessible,
also numerical simulations can be performed in a more controlled way
because typical problems with simulations of heteroclinic switching,
e.g. exponentially increasing switching times and exponentially decreasing
distances to saddles, do not occur if the heteroclinic switching is
replaced by networks of unstable attractors. 

We thank S. Stolzenberg for help during project initiation. Supported
by the Federal Ministry of Education \& Research (BMBF) Germany, Grant
No.~01GQ0430.

\end{document}